\documentclass[11pt,a4paper,fleqn]{scrartcl}
\usepackage{amssymb}
\usepackage{graphicx}
\usepackage{amsmath}
\usepackage{lineno}
\usepackage{xcolor}
\usepackage[normalem]{ulem}

\begin{document}

\title{Changes in behaviour when adherers to an intervention experience a different epidemic than non-adherers}

\author{
	\small{
	Yuan Liu\(^{1,2}\),
	Michael Sieber\(^{2}\),
	Bin Wu\(^{1,*}\),
	Arne Traulsen\(^{2,*}\)
	}
}
\date{
	\footnotesize{
		\(^1\) School of Mathematical Sciences, Beijing University of Posts and Telecommunications, Beijing, China\\
		\(^2\) Department of Theoretical Biology, Max Planck Institute for Evolutionary Biology, Plön, Germany\\
		\(^*\) Bin Wu: bin.wu@bupt.edu.cn, Arne Traulsen: traulsen@evolbio.mpg.de
	}
}

\maketitle

\begin{abstract}
Non-pharmaceutical interventions (NPIs), including mask-wearing, physical distancing, and hygiene measures, provide the primary means of reducing transmission in the early stages of an epidemic. Individuals adopt one of two strategies—adherence (A) or non-adherence (N) to NPIs. These strategies influence the transmission rate and thus the number of infections, but they also come with inherent costs and benefits. We propose a model coupling behavior and disease dynamics in adherers and non-adherers based on the SIR framework.  This gives rise to six behavioral-epidemiological compartments. Using numerical simulations and analytical considerations, we first examine the case where strategies are fixed.  Stronger NPIs and more initial adherers lead to fewer infections, and adherers consistently experience lower infection risk than non-adherers. We then introduce behavioral switching based on the benefits and costs of the two strategies. When NPIs are effective, higher transmission rates promote adherence, resulting in fewer infections. Strikingly, in high-severity outbreaks, even modestly effective NPIs can significantly reduce infections. These findings highlight the critical role of the coupling between behavior and disease dynamics, and underscore how individual choices can compromise or compensate public health interventions.
\end{abstract}

\section{Introduction}
The emergence of infectious diseases such as COVID-19 highlights the persistent threat that epidemics pose to global health \cite{steiner2024}.
Pharmaceutical interventions (PIs), such as vaccines and antiviral treatments, have proven essential in mitigating the impacts of the epidemic \cite{yamana2023, wong2024}.
The impact of reductions in transmission due to non-pharmaceutical interventions (NPIs) has been included in epidemiological models for medium- and long-term scenarios of COVID-19 and other respiratory infections \cite{baker2022long}.
Many previous models combined individual vaccination behavior with disease transmission, yielding fruitful results through evolutionary game theory \cite{bauch2004,perisic2009,wang2016}.
On the other hand, non-pharmaceutical interventions (NPIs), including mask-wearing, physical distancing, and hygiene measures, provide the primary means of reducing transmission in the early stages of an epidemic \cite{reluga:PLOSCB:2010,flaxman2020estimating, mendez2021systematic, lai2020effect}.
But the effectiveness of NPIs largely depends on individual behavior.
Personal decisions regarding whether or not to adhere to NPIs can determine the trajectory of an epidemic, influencing both the pace of transmission and the eventual scale of the outbreak \cite{wei2023adoption, zamir2020non}.

Despite this behavioral diversity, epidemiological models often rely on a simplifying assumption: that individuals in a population behave homogeneously \cite{fenichel2011adaptive, bansal2007individual}.
But individuals re-assess their perceived risk, weigh the costs and benefits of precautionary actions, and adjust their behavior over time \cite{bergstrom:PNAS:2023, liu2022coevolution}.
Some may consistently comply with public health guidelines, while others may ignore them entirely—or change their decisions as the epidemic unfolds. 
Although recent studies have begun to incorporate such behavioral changes into epidemic models, many assume that adherers and non-adherers are equally exposed to infection, or that individuals switch strategies continuously and probabilistically \cite{reluga:PLOSCB:2010,saad-roy:PNAS:2023,shi:MB:2024}. 
Such models capture general behavior-epidemic feedback, but they overlook how behavioral heterogeneity can shape different epidemics within subgroups \cite{bauch2005imitation, bauch2013social, glaubitz2023population, hu2024evolutionary, morsky2023impact}.

Previous work has considered the interaction of behavioral changes to an ongoing epidemic and the epidemiological dynamics \cite{saad-roy:PNAS:2023, karlsson:SciRep:2020, tanimoto:book:2021, morsky2023impact,traulsen:PNAS:2023, flores:npjC:2025, glaubitz:PNAS:2024}. 
These frameworks have provided insights into feedback mechanisms between behavior and disease prevalence, including thresholds for collective behaviors, oscillatory dynamics, and social dilemmas.
Some studies have incorporated differential risk exposure between subpopulations, such as adherers and non-adherers \cite{fenichel2011adaptive, chen2011public, espinoza2022heterogeneous, funk2009spread}. 
However, they generally rely on simplifying assumptions about infection processes or behavioral responses, which limit their ability to fully capture the distinct infection trajectories and feedback effects of these groups.

Here, we introduce a behavioral-epidemiological model that explicitly distinguishes between individuals who adhere to NPIs and those who do not, while allowing for different infection dynamics across these two subpopulations. 
It partitions the population into six compartments, representing both health status (susceptible, infected, recovered) and behavioral strategy (adhering or not adhering). 
Strategy switching is modeled as payoff-driven behavior \cite{hofbauer:book:1998,cressman2014replicator, schuster1983replicator}, influenced by perceived risk, the effectiveness and cost of NPIs, and social learning dynamics. 
By jointly modeling behavioral adaptation and epidemic progression, we aim to uncover the conditions under which voluntary adherence can suppress disease spread—and when it may fall short. 
This framework offers insights into how public health measures can be better tailored to account for individual decision-making, ultimately improving epidemic control.
We aim to better understand the role of adherence in controlling epidemics and how public health measures can be improved to manage disease outbreaks more effectively.

\section{Model}\label{sec2}

\begin{figure}[h]
	\begin{minipage}{0.5\linewidth} 
	\center
	\includegraphics[width=1.1\linewidth]{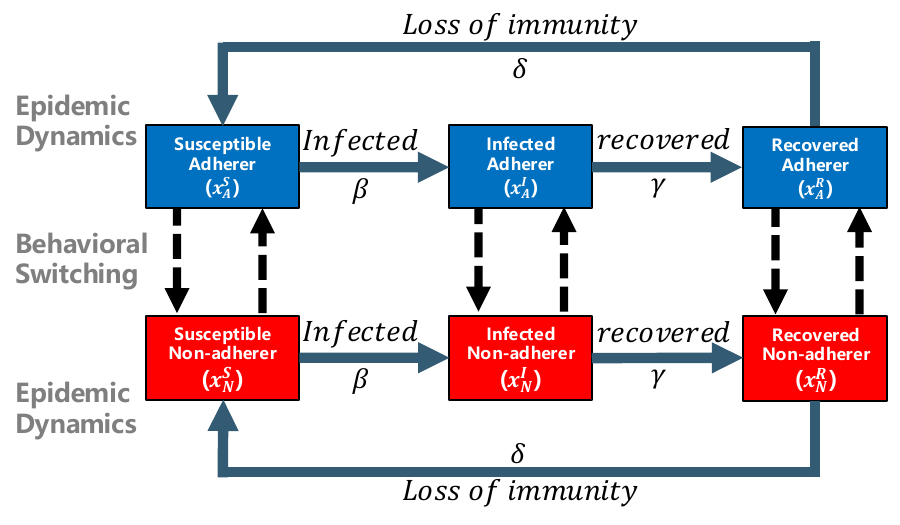}
	\end{minipage}
	\begin{minipage}{0.5\linewidth} 
	\center
	\includegraphics[width=0.7\linewidth]{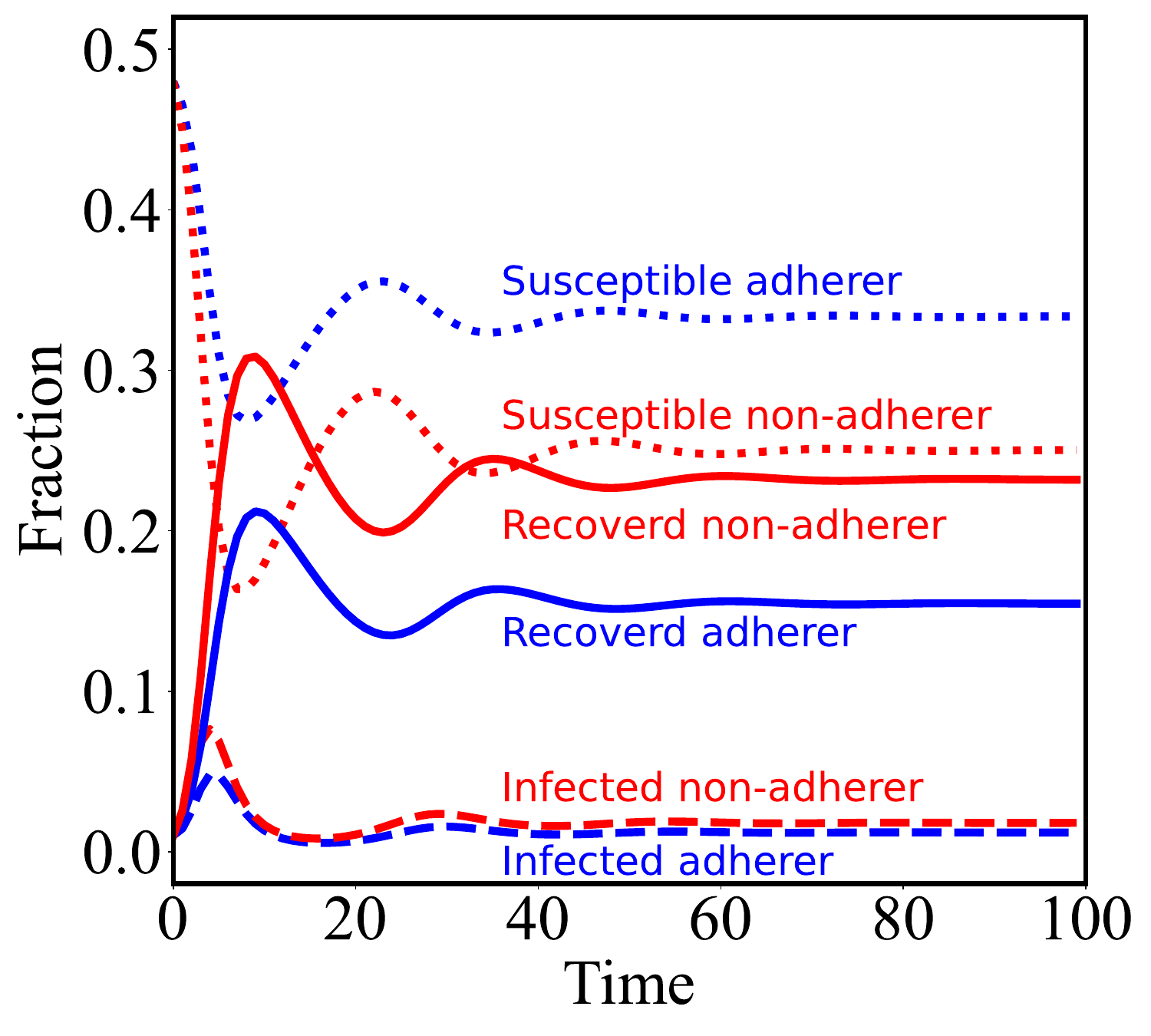}
	\end{minipage}
	\caption{\label{model} \textbf{Schematics and dynamics of the model.} 
	Individuals can choose one of two strategies: adherence (\(A\)), where individuals follow an NPI, or non-adherence (\(N\)), where individuals do not follow this measure. Within each group, individuals are categorized into three health states: susceptible (\(x_A^S\), \(x_N^S\)), infected (\(x_A^I\), \(x_N^I\)), and recovered (\(x_A^R\), \(x_N^R\)).}
\end{figure}

To study how adherence to non-pharmaceutical interventions (NPIs) affects epidemic dynamics, we assume a large population without spatial structure. 
In this model, individuals can choose one of two strategies: adherence (\(A\)), where individuals follow an NPI, or non-adherence (\(N\)), where individuals do not follow this measure. 
Within each group, individuals are categorized into three health states: susceptible (with proportions \(x_A^S\), \(x_N^S\)), infected (\(x_A^I\), \(x_N^I\)), and recovered (\(x_A^R\), \(x_N^R\)) (see Fig.\ref{model}). 

In the epidemic processes of infection, recovery, and loss of immunity, individuals change only their health states (\(S\), \(I\), \(R\)) without altering their strategies (\(A\), \(N\)).
The proportions of adherers and non-adherers are represented by \(x_A\) and \(x_N\),
where \(x_A = x_A^S +x_A^I +x_A^R\), \(x_N= x_N^S +x_N^I +x_N^R\), and \(x_A + x_N = 1\). 

The transmission rate for susceptible individuals (\(x_A^S\) and \(x_N^S\)) when they encounter infected individuals (\(x_A^I\) and \(x_N^I\)) is denoted by \(\beta\). 
NPIs reduce the transmission rate, their effectivity is given by \(e\), such that the probability of transmission is reduced to \(1-e\).
When the susceptible individual meets the infected, either of them following the NPIs play the same role in the disease transmission.
In other words, the effective transmission rate becomes \( (1-e) \beta\) if either of the two individuals in an interaction is following the intervention.
This reduction is assumed to be the same whether the infected individual or the uninfected follows the NPI \cite{saad-roy:PNAS:2023,flores:npjC:2025}-- this assumption reduces the number of parameters substantially and allows to recover the parameter regimes discussed in \cite{traulsen:PNAS:2023} for more general settings.
In addition, we include a birth and death at a rate \(\mu\) to include an influx of susceptible individuals. 
Infected individuals (\(x_A^I\), \(x_N^I\)) recover and enter the recovered state (\(x_A^R\), \(x_N^R\)) after an average recovery time of \(1/\gamma\).
After a time of \(1/\delta\), individuals in the recovered state become susceptible again, which reflects the loss of immunity over time.

The payoff for interactions between individuals depends on their strategies. 
When a non-adherer interacts with another non-adherer, the payoff is \(-\beta \xi x_N^I \). 
Here, \( \xi \geq 0\) represents the perceived risk of becoming infected rather than the biological transmission process itself. 
It captures how individuals evaluate infection risk and disease severity when making behavioral decisions, and therefore influences the payoff structure governing behavioral switching. 
When a susceptible non-adherer interacts with an infected adherer, the payoff is \(- (1 - e) \beta \xi  \). 
When a susceptible non-adherer interacts with an infected non-adherer, the payoff is \(- \beta \xi  \). 
In interactions with susceptible or recovered individuals, the payoff of the non-adherer is $0$.
 The average payoff for non-adherers is then
\[
\pi_{N} = - (1 - e) \beta \xi  x_A^I - \beta \xi x_N^I.
\]
For adherers, following NPIs comes with a cost \(c\), representing time, resources, or effort. 
When a susceptible adherer interacts with an infected non-adherer, the payoff is \( - (1 - e) \beta \xi - c \), and when interacting with an infected adherer, the payoff is \(- (1 - e)^2 \beta \xi - c \). 
In other interactions, the payoff of the non-adherer is $-c$.
Thus, the average payoff for adherers is
\[
\pi_{A} = - (1 - e)^2 \beta \xi x_A^I - (1 - e) \beta \xi x_N^I  - c.
\]
Thus, the payoff matrix is given by:
\begin{equation*}
\label{game1}
\bordermatrix{
     &  A         &  N \cr
 A & - (1 - e)^2 \beta \xi x_A^I  - c  &- (1 - e) \beta \xi x_N^I   - c  \cr
 N & - (1 - e) \beta \xi x_A^I & - \beta \xi x_N^I 
 }.
\end{equation*}

To model how individuals change their strategies, we assume that they switch strategies based on the differences in their payoffs. The rate at which they switch strategies is proportional to the payoff differences, following the replicator dynamics \cite{hofbauer:book:1998}.
E.g. the temporal change in the number of susceptible adherers from behavioral changes is \( x_N^S x_A (\pi_{A}\!-\!\pi_{N}) - x_A^S x_N(\pi_{N}\!-\!\pi_{A}) \).
The overall dynamics is given by
\begin{align}
\label{dynamics3}
\begin{split}
\tfrac{d}{dt}{x_A^S} =& \underbrace{- (1-e) \beta((1-e) x_A^I + x_N^I)  x_A^S  + \delta x_A^R}_{\text epidemiology}  + \underbrace{\mu x_A - \mu x_A^S}_{\text birth \& death}\\
& + \underbrace{\tfrac{1}{\tau} [x_N^S x_A (\pi_{A}\!-\!\pi_{N}) - x_A^S x_N(\pi_{N}\!-\!\pi_{A})]}_{\text behavior}\\
\tfrac{d}{dt}{x_N^S} =&- \beta( (1-e) x_A^I + x_N^I)  x_N^S + \delta x_N^R  +\mu x_N - \mu x_N^S \\
&  - \tfrac{1}{\tau} [x_N^S x_A (\pi_{A}\!-\!\pi_{N}) - x_A^S x_N(\pi_{N}\!-\!\pi_{A})]\\
\tfrac{d}{dt}{x_A^I} =& + (1-e) \beta( (1-e) x_A^I + x_N^I)  x_A^S - \gamma x_A^I  - \mu x_A^I \\
& + \tfrac{1}{\tau} [x_N^I x_A (\pi_{A}\!-\!\pi_{N}) - x_A^I x_N(\pi_{N}\!-\!\pi_{A})]\\
\tfrac{d}{dt}{x_N^I} =&  + \beta((1-e) x_A^I + x_N^I)  x_N^S - \gamma x_N^I  - \mu x_N^I\\
& - \tfrac{1}{\tau} [x_N^I x_A (\pi_{A}\!-\!\pi_{N}) - x_A^I x_N(\pi_{N}\!-\!\pi_{A})]\\
\tfrac{d}{dt}{x_A^R} =&+ \gamma x_A^I - \delta x_A^R   - \mu x_A^R \\
& + \tfrac{1}{\tau} [x_N^R x_A (\pi_{A}\!-\!\pi_{N}) - x_A^R x_N(\pi_{N}\!-\!\pi_{A})]\\
\tfrac{d}{dt}{x_N^R} =& + \gamma x_N^I - \delta x_N^R  - \mu x_N^R \\
& - \tfrac{1}{\tau} [x_N^R x_A (\pi_{A}\!-\!\pi_{N}) - x_A^R x_N(\pi_{N}\!-\!\pi_{A})]
\end{split}
\end{align}
Here, \(\tau\) represents the relative time scale of behavioral changes compared to the epidemiological dynamics.
If \(\tau > 1\), the epidemic dynamics is faster than the switching between strategies. 
If \(0< \tau < 1\), switching between strategies is faster. 
If \(\tau \to +\infty\), the epidemic dynamics is much faster than the switching between strategies; in this case, we could ignore the impact of switching between strategies.

\section{Results}\label{sec3}
\subsection{Epidemic Dynamics With Behavioral Switching}\label{subsec2}

\begin{figure}
	\begin{minipage}{1\linewidth} 
	\includegraphics[width=1\linewidth]{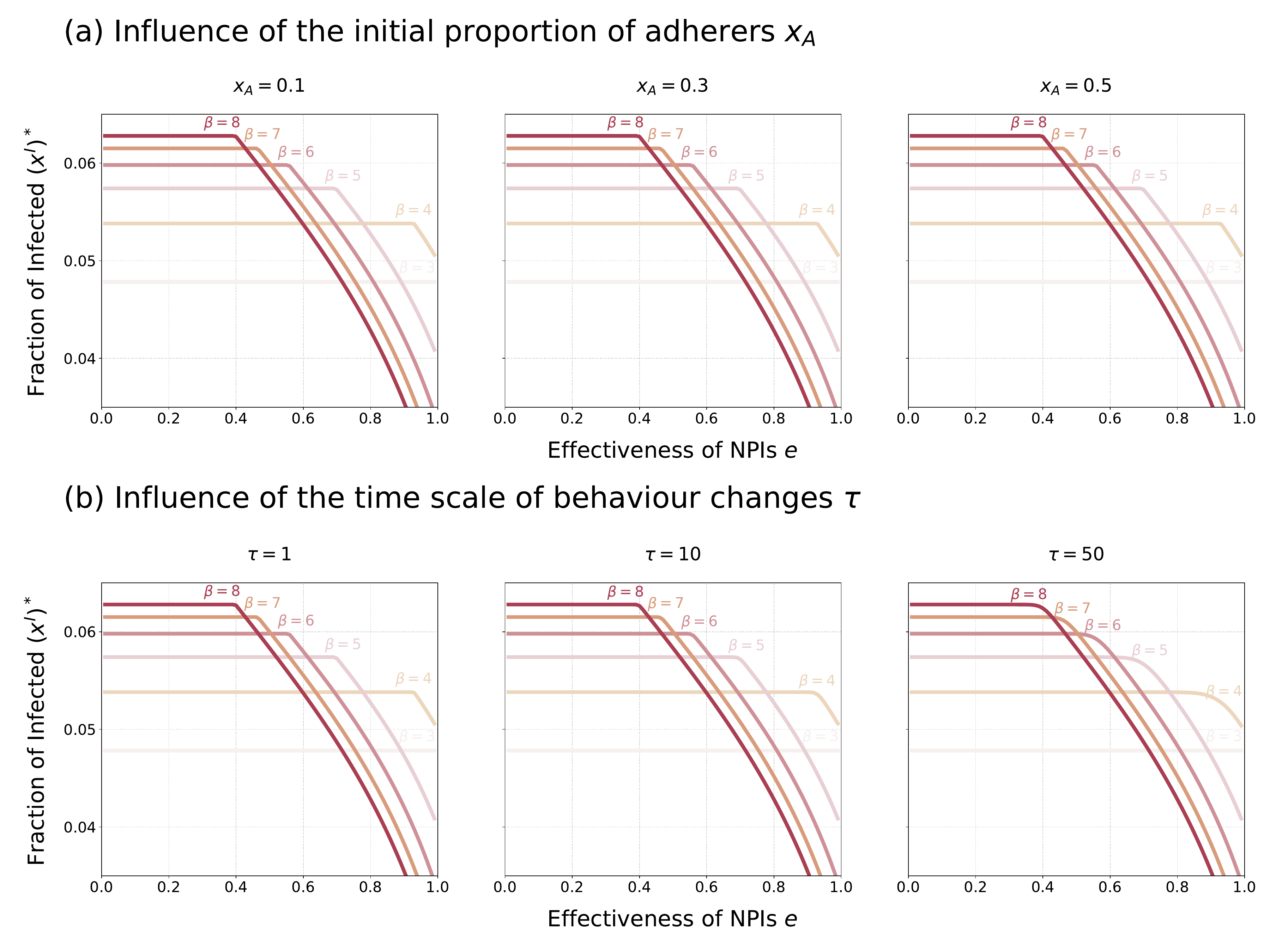}
	\end{minipage}
	\caption{\label{fig3-1} 
	\textbf{Infections at equilibrium.} 
	(a) Fraction of infected individuals at equilibrium ($x_A^I + x_N^I$) as a function of the effectiveness of NPIs 
$e$, for different initial proportions of adherers $x_A = 0.1, 0.3, 0.5$ and transmission rates $\beta = 3, 4, 5, 6, 7, 8$, based on the dynamics given by Eq.\ref{dynamics3}. 
	(b) Fraction of infected individuals at equilibrium ($x_A^I + x_N^I$) as a function of the effectiveness of NPIs 
$e$, for different time scales $\tau= 1, 10, 50$ and transmission rates $\beta = 3, 4, 5, 6, 7, 8$
	(Parameters: $\mu = \tfrac{1}{50 \cdot 52}, \gamma = 1, \delta = \tfrac{1}{0.25  \cdot 52}, c = 1, \xi = 5, \tau = 1$).}
\end{figure}

Now, we consider that the epidemic dynamics and the switching between strategies occur on comparable time scales, i.e.,  $\tau \approx 1$ 
in Eq.~\eqref{dynamics3}.
We analyze the proportion of infected individuals at equilibrium under varying levels of transmission rate \(\beta\), NPI effectiveness \(e\), initial adherence \(T_A\)  and time scale $\tau$ (Fig.\ref{fig3-1}). 
A higher initial fraction of individuals adhering to NPIs lead to a slightly lower equilibrium level of infection, but the equilibrium is largely unaffected
by the initial condition.
As the transmission rate \(\beta\) increases, the infection burden becomes higher, even when NPIs are moderately effective. 
Conversely, increasing the effectiveness of NPIs reduces the fraction of infected individuals across all values of \(\beta\). 
Notably, when \(e\) approaches \(1\), the infection levels drop sharply, even for high transmission rates. 
We further examine the role of time scale \(\tau\), which represents the relative time scale of behavioral switching 
compared to epidemic dynamics. 
Interestingly, the infection curves are only slightly affected by the value of \(\tau\). 
This suggests that the final size of the outbreak is more sensitive to transmission rates and NPI effectiveness than to delays in behavioral switching.

\begin{figure}
	\centering
	\begin{minipage}{0.68\linewidth} 
	\includegraphics[width=1\linewidth]{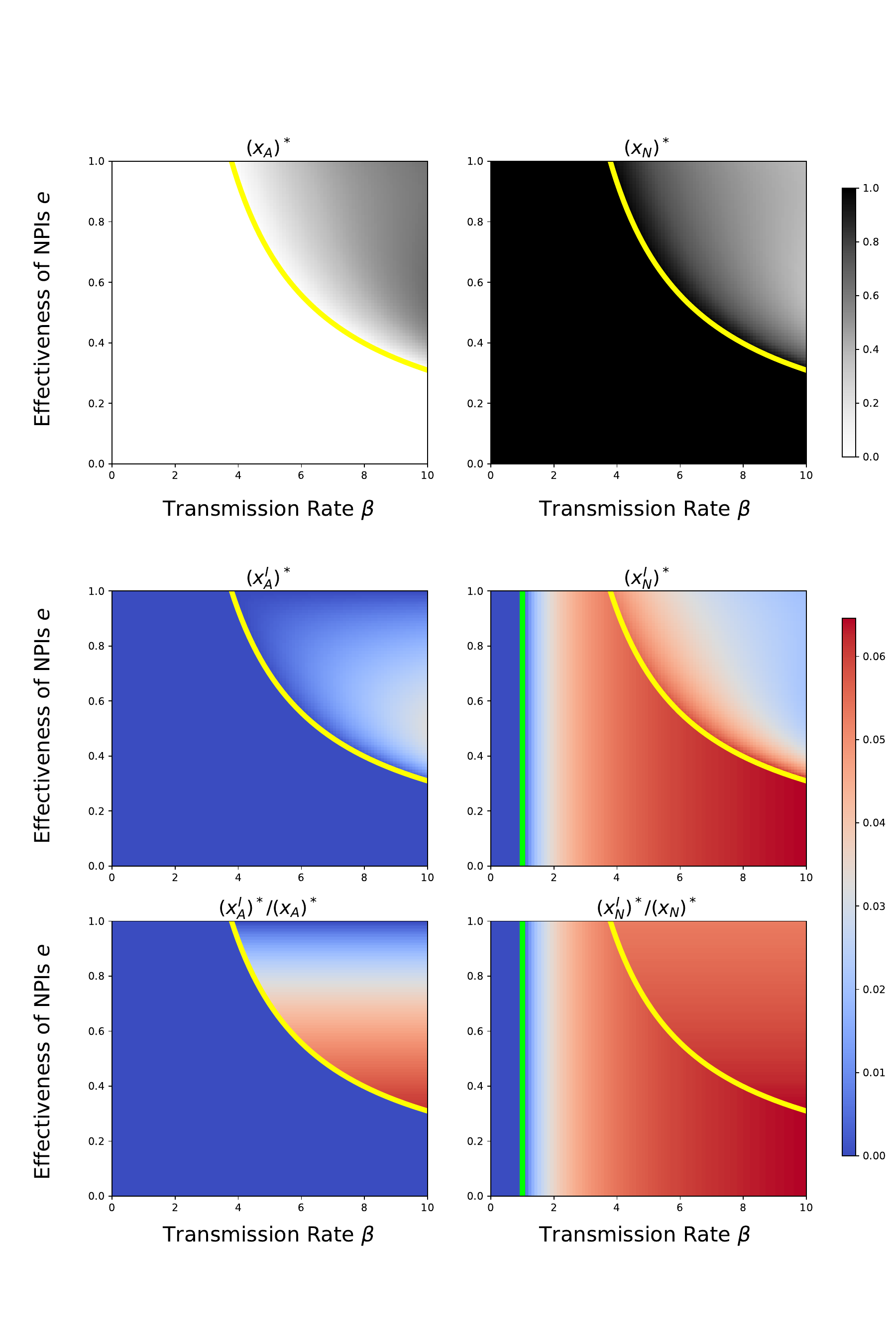}
	\end{minipage}
	\caption{\label{fig3-4} 
	\textbf{Qualitative changes of the infection dynamics with the transmission rate.} 
	The yellow line denotes the adherence invasion boundary, representing the condition under which adherers can start to establish within the population.
	The green line shows that for \(R_0=\frac{\beta}{\gamma}<1\), there are no infections. 
	(a) Total proportion of adherers and non-adherers at equilibrium based on the dynamics given by Eq.\ref{dynamics3}.
	(b) Fraction of infected individuals in two groups (adherers and non-adherers) at equilibrium based on the dynamics given by Eq.\ref{dynamics3}.
	It shows the final proportion of infected individuals in both groups across a range of transmission rate ($\beta$) and effectiveness of NPIs ($e$). 
	The top-left panel displays the infected adherents ($({x_A^I})^*$), the top-right panel shows the infected non-adherents ($({x_N^I})^*$), the bottom-left panel represents the proportion of infected adherents in the total adherent population ($({x_A^I})^*/(x_A)^*$), and the bottom-right panel represents the proportion of infected non-adherents in the total non-adherent population ($({x_N^I})^*/(x_N)^*$). 
	Parameters: $\mu = \tfrac{1}{50 \cdot 52}, \gamma = 1, \delta = \tfrac{1}{0.25  \cdot 52}, c = 1, \xi = 5, \tau=1$, initial adherent proportion $x_A = 0.5$.}
\end{figure}

\begin{figure}
	\begin{minipage}{1\linewidth} 
	\includegraphics[width=1\linewidth]{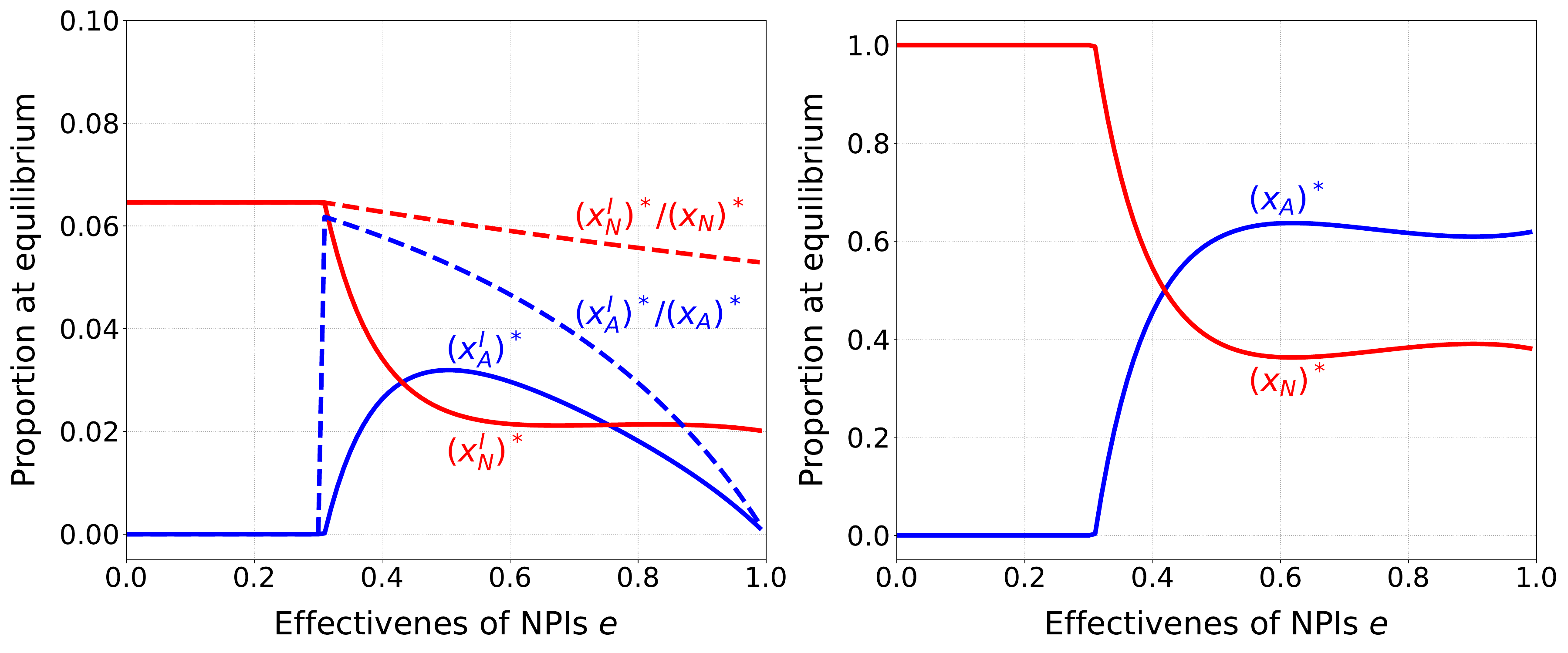}
	\end{minipage}
	\caption{\label{fig3-5} 
	\textbf{Adherence is maximised for intermediate effectiveness.} 
	(a) Infected adherers ($({x_A^I})^*$), infected non-adherers ($({x_N^I})^*$), and proportion of infected individuals in both adherent and non-adherent populations at equilibrium, based on the dynamics given by Eq.\ref{dynamics3}. 
	(b) Total proportion of adherers and non-adherers at equilibrium based on the dynamics given by Eq.\ref{dynamics3}.
	Parameters: $\beta = 10, \mu = \tfrac{1}{50 \cdot 52}, \gamma = 1, \delta = \tfrac{1}{0.25  \cdot 52}, c = 1, \xi = 5, \tau=1$, initial adherent proportion $x_A = 0.5$.}
\end{figure}

\begin{figure}
	\centering
	\begin{minipage}{0.7\linewidth} 
	\includegraphics[width=1\linewidth]{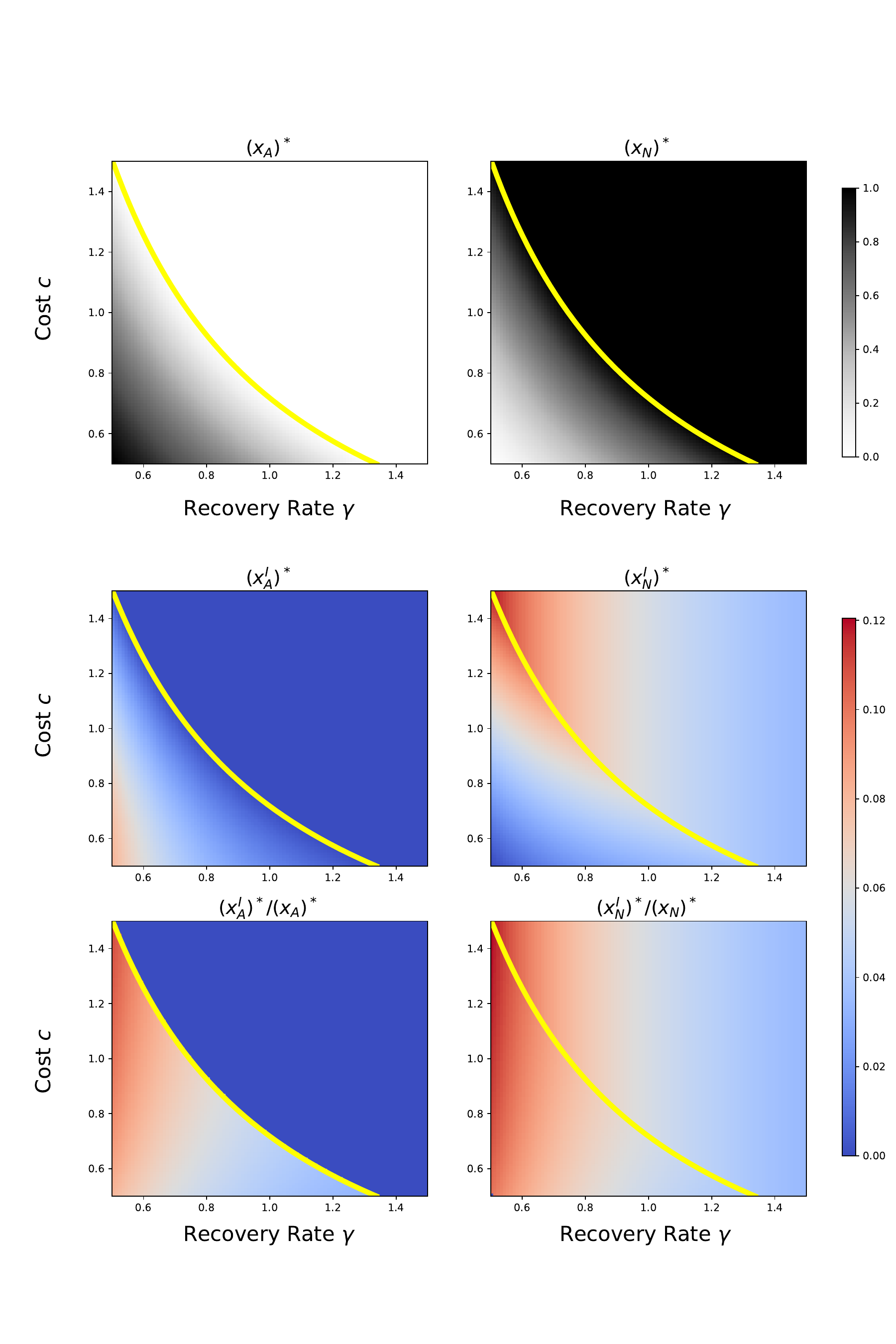}
	\end{minipage}
	\caption{\label{fig3-6} 
		\textbf{Qualitative changes of the infection dynamics with the recovery rate.} 
	The yellow line denotes the invasion boundary, representing the condition under which adherers can start to establish within the population.
	(a) Total proportion of adherers and non-adherers at equilibrium based on the dynamics given by Eq.\ref{dynamics3}.
	(b) Fraction of infected individuals in two groups (adherers and non-adherers) at equilibrium based on the dynamics given by Eq.\ref{dynamics3}.
	It shows the final proportion of infected individuals in both groups across a range of recovery rate ($\gamma$) and the cost of adherence to follow NPIs ($c$). 
	The top-left panel displays the infected adherents ($({x_A^I})^*$), the top-right panel shows the infected non-adherents ($({x_N^I})^*$), the bottom-left panel represents the proportion of infected adherents in the total adherent population ($({x_A^I})^*/(x_A)^*$), and the bottom-right panel represents the proportion of infected non-adherents in the total non-adherent population ($({x_N^I})^*/(x_N)^*$). Parameters: $\mu = \tfrac{1}{50 \cdot 52}, e = 0.5, \beta=5, \delta = \tfrac{1}{0.25  \cdot 52}, \xi = 5, \tau=1$, initial adherent proportion $x_A = 0.5$.}
\end{figure}

\begin{figure}
	\centering
	\begin{minipage}{0.7\linewidth} 
	\includegraphics[width=1\linewidth]{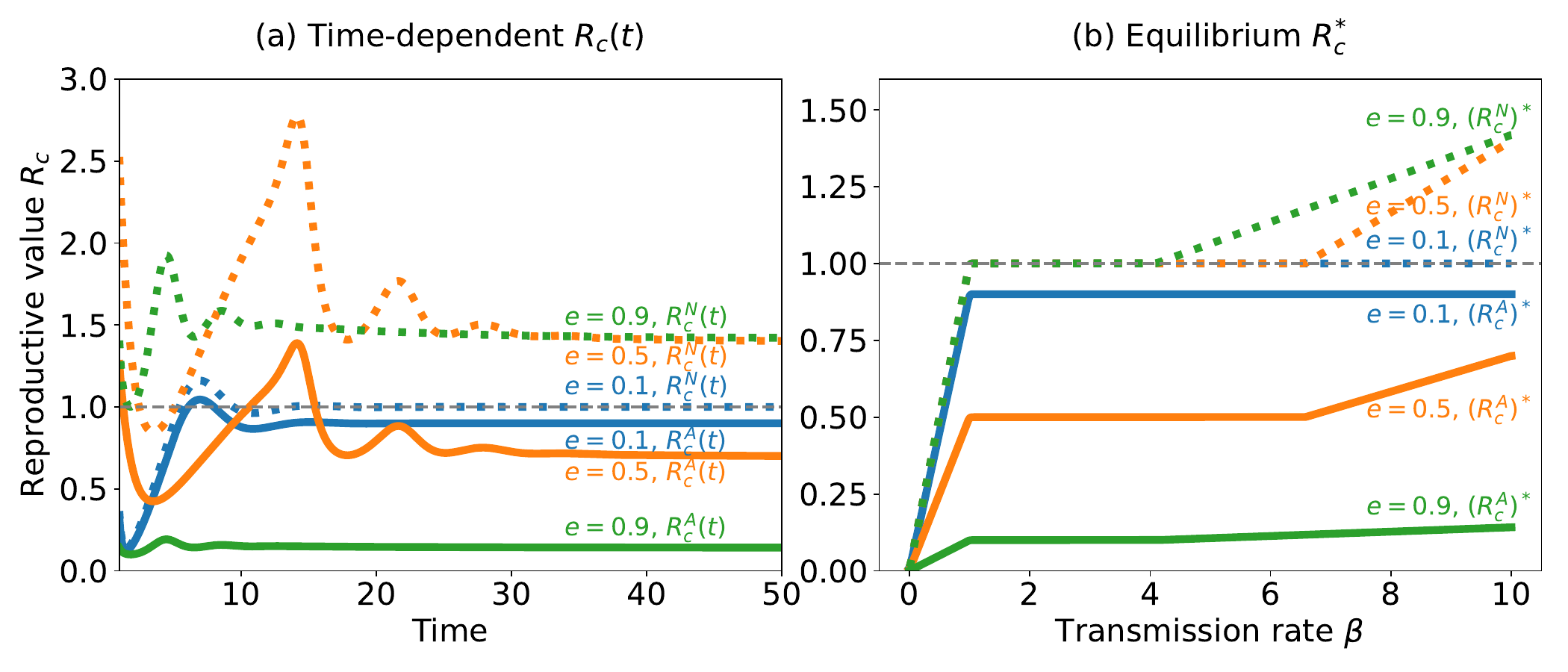}
	\end{minipage}
	\caption{\label{reproduction} 
	\textbf{Temporal dynamics of the infection dynamics.} 
	(a) Dynamics of the effective reproduction values for adherers, $R_c^A =\frac{\beta}{\gamma} (1 - e) ((1 - e) x_A^S + x_N^S )$,  
	and for non-adherers, $R_c^N =\frac{\beta}{\gamma} ((1 - e) x_A^S + x_N^S )$, for $\beta = 10$.
	(b) Effective reproduction values at equilibrium ${R_c^A}^*$ and ${R_c^N}^*$ 
	(Parameters: 
	$\mu = \tfrac{1}{50 \cdot 52}, \gamma=1,  \delta = \tfrac{1}{0.25  \cdot 52}, \xi = 5, \tau=1$).}
\end{figure}

We consider the impact of varying both the transmission rate ($\beta$) and the effectiveness of NPIs ($e$) on the infection level when the initial proportion of adherers and non-adherers is 0.5 (see Fig.\ref{fig3-4}).
We show the invasion boundary, \(e<  \frac{c}{\beta \xi (x_N^I)*} \), representing the condition under which adherers can establish within the population (see Appendix \ref{appendix2}).
For \(e<  \frac{c}{\beta \xi (x_N^I)*} \), we expect no adherers in the population.
When the disease spreads rapidly (i.e., $\beta$ is large), the proportion of infected adherers increases. 
It decreases as the effectiveness of NPIs ($e$) rises (see Fig.\ref{fig3-5}). 
At the same time, the proportion of infected non-adherers decreases first and then increases.
We also calculate the effective reproduction values for non-adherers, $R_c^N =\frac{\beta}{\gamma} ((1 - e) x_A^S + x_N^S )$, and adherers, $R_c^A = (1-e) R_c^N$. These are the average number of secondary infections in both groups caused by a single infectious individual during its infectious period in a wholly susceptible population (see Fig. \ref{reproduction}).
Increasing the effectiveness of NPIs $e$, $R_c^N$ reduces, while $R_c^A$ is first increasing and then decreasing.
Similarly, we obtain the impact of varying both the recovery rate ($\gamma$) and the cost of adherence to follow NPIs ($c$) on the infection level when the initial proportion of adherers and non-adherers is 0.5 (see Fig.\ref{fig3-6}).
The fraction of infected increases if the recovery rate $\gamma$ is low and the the cost $c$ is large.
For the invasion boundary \(c>  e \beta \xi (x_N^I)* \), no individuals follow the NPIs.

\subsection{Epidemic Dynamics Without Behavioral Switching}\label{subsec1}

\begin{figure}
	\begin{minipage}{1\linewidth} 
	\includegraphics[width=1\linewidth]{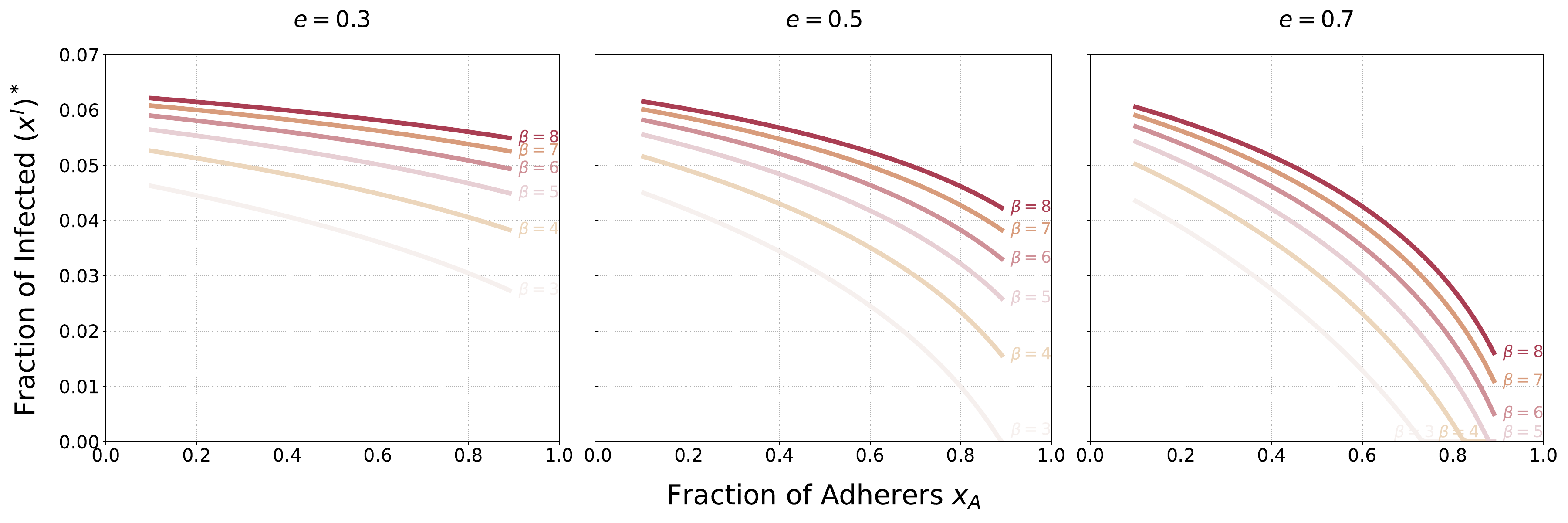}
	\end{minipage}
	\caption{\label{fig1-1} 
		\textbf{The fraction of infected individuals decreases with constant fraction of adherers and the effectiveness.} 
	Fraction of infected individuals at equilibrium for varying the effectiveness of NPIs ($e$), the proportion of adherers ($x_A$) and the transmission rate ($\beta$), based on the dynamics given by Eq.\ref{dynamics1}.
	We show the fraction of infected individuals ($x_A^I + x_N^I$) for each $x_A  (0.1, 0.3, 0.5)$ across six different values of $\beta (3, 4, 5, 6, 7, 8)$ 
		(Parameters: $\mu = \tfrac{1}{50 \cdot 52}, \gamma = 1, \delta = \tfrac{1}{0.25  \cdot 52}$).
	}
\end{figure}

Next, we assume that the epidemic dynamics is much faster than the switching between strategies, i.e.,  $\tau \to + \infty$.
Such an assumption is supported by real-world examples of fast-spreading infectious diseases. 
For instance, seasonal influenza and measles can transmit rapidly within days, while individual behavioral adaptations, such as adherence to non-pharmaceutical interventions, may occur on longer timescales. 
Similarly, in the early stages of SARS-CoV-2 outbreaks, case numbers increased exponentially before widespread behavioral or policy responses were implemented.
In this case, the dynamics reduces to 
\begin{equation}
\label{dynamics1}
\begin{split}
\tfrac{d}{dt}{x_A^S} =& - (1-e) \beta( (1-e) x_A^I + x_N^I)  x_A^S + \delta x_A^R+ \mu x_A- \mu x_A^S\\
\tfrac{d}{dt}{x_N^S} =& - \beta( (1-e) x_A^I + x_N^I)  x_N^S  + \delta x_N^R +\mu x_N - \mu x_N^S \\
\tfrac{d}{dt}{x_A^I} =&+(1-e) \beta((1-e) x_A^I + x_N^I)  x_A^S - \gamma x_A^I- \mu x_A^I \\
\tfrac{d}{dt}{x_N^I} =& +\beta((1-e) x_A^I + x_N^I)  x_N^S - \gamma x_N^I - \mu x_N^I \\
\tfrac{d}{dt}{x_A^R} =&\gamma x_A^I  - \delta x_A^R -\mu x_A^R \\
\tfrac{d}{dt}{x_N^R} =&\gamma x_N^I - \delta x_N^R - \mu x_N^R.
\end{split}
\end{equation}
To infer when the separation between disease status of adherers and non-adherers makes a substantial difference, we focus on the proportion of total infected individuals at the endemic equilibrium (see Fig.\ref{fig1-1}). 
In our case, the basic reproductive number \(R_0\) is given by \(R_0=\frac{\beta}{\gamma+ \mu}\) \cite{heffernan:JRSI:2005,traulsen:NAL:2022}. 
For \(R_0<1 \), infections do not spread.
For \(R_0>1 \), the proportion of infected individuals at equilibrium is influenced by the constant proportion of adherers ($x_A$) for various values of the transmission rate ($\beta$) and the effectiveness of NPIs ($e$), based on the dynamics given by Eq.\ref{dynamics1}. 
As the proportion of adherers increases, the equilibrium proportion of infected individuals decreases. 
Furthermore, an increase in the transmission rate $\beta$ leads to a higher proportion of infected individuals. 
If the effectiveness of NPIs $e$ decreases, the proportion of infected individuals increases.
In other words, for more transmissible diseases and less effective interventions, individuals are more likely to become infected.

\begin{figure}
	\begin{minipage}{1\linewidth} 
	\includegraphics[width=1\linewidth]{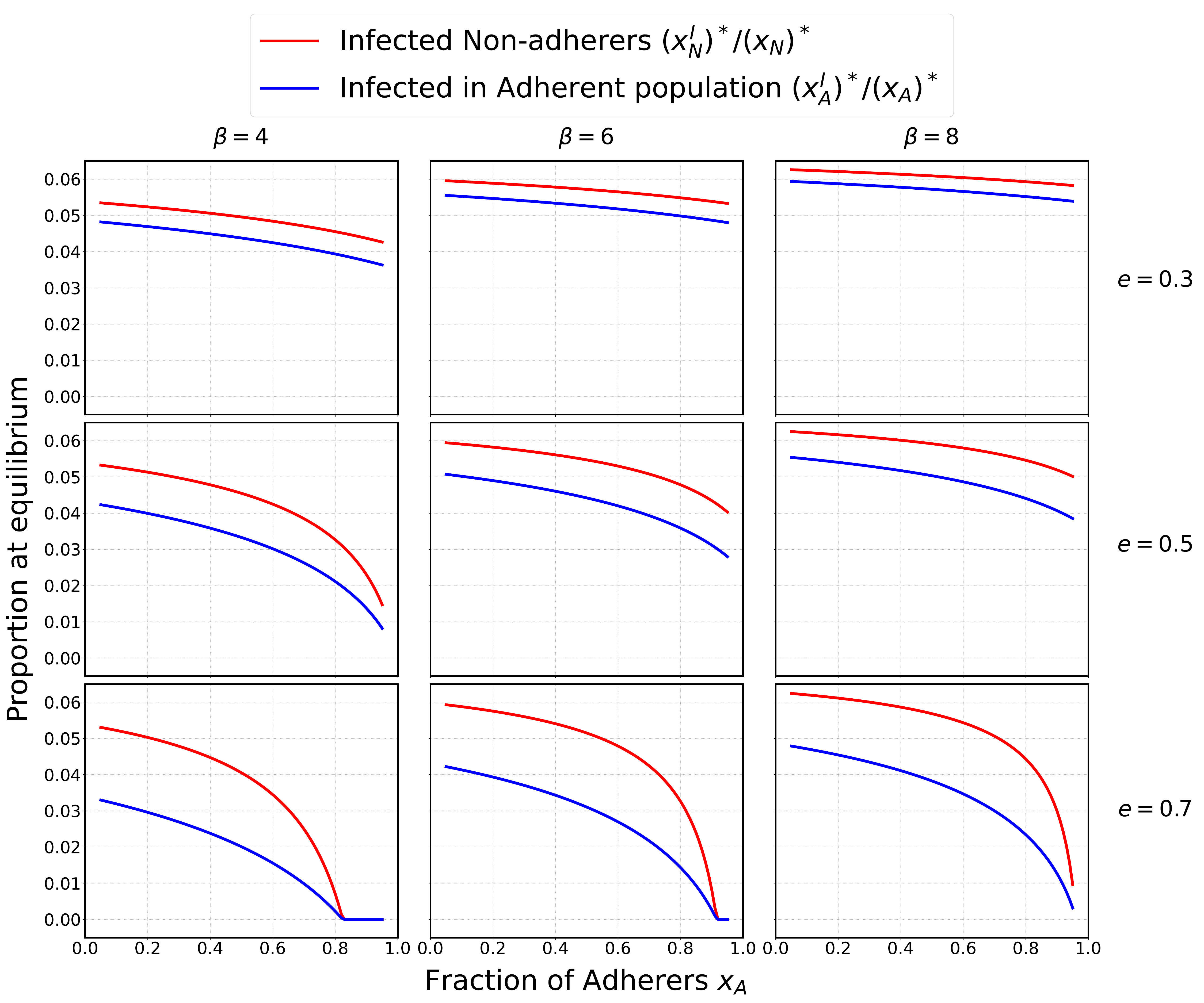}
	\end{minipage}
	\caption{\label{fig1-4} 
	\textbf{Increasing the fraction of adherers reduces infections in both non-adherers and adherers.} 
	Fraction of infected individuals in both adherent and non-adherent populations at equilibrium as a function of the fraction of adherers($x_A$) for various values of the transmission rate ($\beta$) and the effectiveness of NPIs ($e$), based on the dynamics given by Eq.\ref{dynamics1}. 
	The blue curve represents the proportion of infected adherents in the total adherent population ($({x_A^I})^*/(x_A)^*$, 
	and the red curve represents the proportion of infected non-adherents in the total non-adherent population ($({x_N^I})^*/(x_N)^*$). 
	For very effective measures (large $e$) and not too large $\beta$, a sufficiently large fraction of adherers can make the number of infections go down to zero
	(Parameters: $\mu = \tfrac{1}{50 \cdot 52}, \gamma = 1, \delta = \tfrac{1}{0.25  \cdot 52}$.)
	}
\end{figure}

Within each strategy group, the fraction of infected individuals among adherers (\(A\)-strategy group) consistently remains lower than that among non-adherers (\(N\)-strategy group), which shows that following NPIs is an effective way for individuals to protect themselves
 (see Fig.\ref{fig1-4}).
 
\begin{figure}
	\begin{minipage}{1\linewidth} 
	\includegraphics[width=1\linewidth]{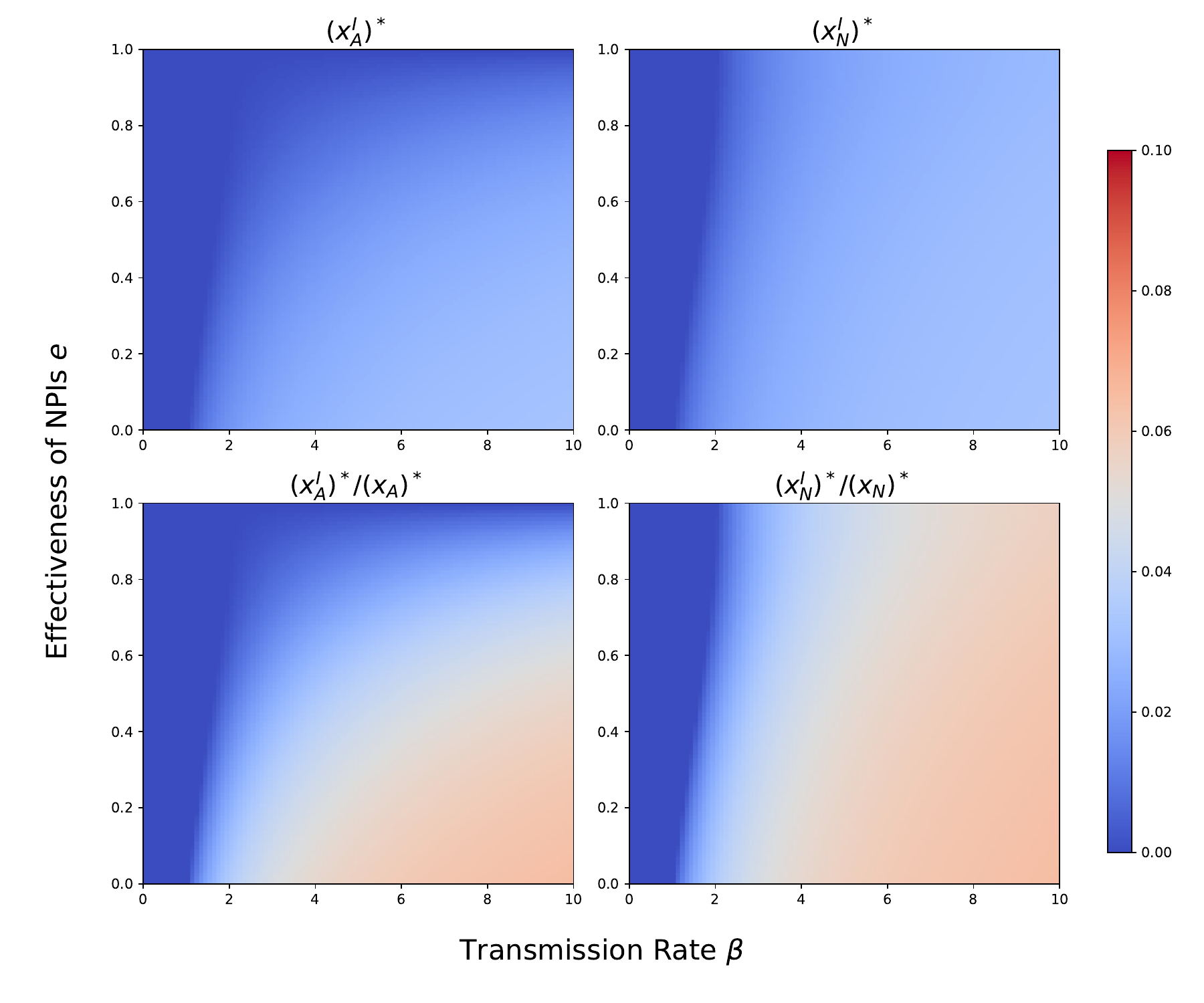}
	\end{minipage}
	\caption{\label{fig1-5} 
	\textbf{Differences between infected adherers and infected non-adherers are most visible when their relative proportion is considered.} 
	Fraction of infection at equilibrium in the two groups (adherers and non-adherers) based on the dynamics given by Eq.\ref{dynamics1}.
	We show the equilibrium proportion of infected individuals in both groups across a range of transmission rate ($\beta$) and effectiveness of NPIs ($e$). 
	For \(R_0=\frac{\beta}{\gamma+\mu}<1 \), there are no infections. 
	The top-left panel displays the infected adherents ($({x_A^I})^*$), the top-right panel shows the infected non-adherents ($({x_N^I})^*$), the bottom-left panel represents the proportion of infected adherents in the total adherent population ($({x_A^I})^*/(x_A)^*$), and the bottom-right panel represents the proportion of infected non-adherents in the total non-adherent population ($({x_N^I})^*/(x_N)^*$)
	(Parameters: $\mu = \tfrac{1}{50 \cdot 52}, \gamma = 1, \delta = \tfrac{1}{0.25  \cdot 52}$, initial adherent proportion $x_A = 0.5$).}
\end{figure}

Finally, we consider the impact of varying both the transmission rate ($\beta$) and the effectiveness of NPIs ($e$) on the infection level when the fraction of adherers and non-adherers is 0.5 (see Fig.\ref{fig1-5}). 
When the disease spreads rapidly (i.e., $\beta$ is large) or the effectiveness of NPIs is low (i.e., $e$ is small), the proportion of infected individuals increases.

\section{Conclusion and Discussion}\label{sec4}
Our model captures the coupled dynamics of epidemic spread and individual behavioral decisions by explicitly distinguishing between adherers and non-adherers to NPIs. 
When assuming that epidemic dynamics is much faster than behavioral switching, we observe that increasing the initial proportion of adherers consistently lowers the equilibrium prevalence of infection. 
This highlights the critical role of early and widespread adherence to NPIs in mitigating disease spread. 
Conversely, increasing the transmission rates or reducing the effectiveness of NPIs elevate infection levels, underscoring the vulnerability of populations when control measures are insufficient or poorly implemented.

Our results reveal persistent heterogeneity in infection risk between adherers and non-adherers: adherers experience much lower infection rates at equilibrium, confirming the personal protective benefits of adherence.
This differentiation goes beyond previous models that assumed homogeneous infection risk across behavioral groups. 
Our framework therefore provides a more nuanced understanding of how individual choices shape epidemic outcomes and feedback into transmission dynamics.

Allowing the timescale of behavioral adaptation to vary, we find that the relative speed of behavioral switching has only a modest effect on equilibrium infection levels compared to the dominant influences of transmission rate and effectiveness of NPIs. 
This suggests that timely implementation and effectiveness of interventions may be more important than the rapidity of behavioral changes for long-term epidemic control. 
Additionally, our analysis of infection prevalence within each behavioral subgroup across parameter regimes reveals complex, non-monotonic patterns: 
For example, as NPIs become more effective, infection among adherers first increases and then decreases, while non-adherers show the opposite trend. 
This can lead to the superficially paradoxical scenario that adherers make up most of the infected cases, especially at intermediate NPI effectiveness (Fig.\ref{fig3-5}a).
But this is of course because the majority of people chose to adhere to NPIs if they are effective enough --- the relative number of infections in this group is always lower than in the group of non-adherers (Fig.\ref{fig3-5}a).
These dynamics point to intricate socio-epidemiological feedbacks that could influence the design of adaptive public health strategies.

In summary, our extended SIR-behavioral model, which separates the population into six compartments based on both health status and adherence to NPIs, provides additional insights into the interplay between individual behavior and epidemic dynamics. 
By explicitly modeling differential infection risks for adherers and non-adherers and allowing strategic switching based on payoff differences, we capture realistic heterogeneity that shapes the trajectory and final size of outbreaks.

A key assumption in our model is that adherence to non-pharmaceutical interventions (NPIs) reduces infection risk equally for both infected and susceptible individuals. 
This simplification affects the payoff structure that governs behavioral switching, and it may influence the resulting coupled behavior-disease dynamics.
For instance, if NPIs were more effective at reducing transmission from infected individuals than protecting susceptibles, the relative advantage of adherence could change, potentially altering thresholds for behavioral adoption and the magnitude of oscillatory dynamics. 
While this assumption allows us to capture the qualitative feedback between behavior and epidemic spread, future work could relax it to examine how differential effectiveness of NPIs modifies epidemic outcomes and collective behavior patterns.

Our findings reinforce the crucial importance of promoting early and sustained adherence to NPIs, especially when vaccines or pharmaceutical interventions are unavailable or limited. 
They also suggest that policies enhancing the effectiveness of NPIs can substantially reduce disease burden, even when individual behavioral responses adjust more slowly than epidemic progression. 
Future work could extend this framework to incorporate vaccination dynamics, heterogeneous contact networks, and more detailed behavioral decision processes, further informing public health responses in ongoing and future epidemics.

\section*{Funding}
This work is supported in part by funds from the National Natural Science Foundation of China (NSFC) under Grant 61751301, the BUPT Excellent Ph.D. Students Foundation under Grant No.CX2023209 and the China Scholarship Council under Grant No.202306470079.
M.S. and A.T. acknowledge generous core funding by the Max Planck Society.

\section*{Author contributions statement}
Y.L., M.S., B.W. and A.T. conceived the study,  
Y.L., A.T. conducted the study, 
Y.L., M.S. and A.T. analyzed the results.  
Y.L. led the writing of the manuscript with significant contributions from all authors.

\section*{Acknowledgement}
A.T. is grateful to Chadi Saad-Roy (UBC Vancouver) for helpful discussions on this subject.

\appendix
\section{The reproductive number.}
In a population without any adherence the fraction of infected individuals \(x_N^I\) will
increase if \( \beta x_N^S - \gamma - \mu >0 \).
If there are almost only susceptible individuals, \(x_N^S \approx 1\),
this is equivalent to \( R_0 = \frac{\beta}{ \gamma +\mu} >1 \), where
$R_0$ is the basic reproduction value  \cite{heffernan:JRSI:2005,traulsen:NAL:2022}. 
For our parametrisation, we have \(R_0\approx\beta\). 

We also calculate the effective reproduction values arising from non-adherers and from adherers separately. 
The cases arising from non-adherers is $R_0$ multiplied by the fraction of susceptibles they can infect, 
$R_c^N =\frac{\beta}{\gamma + \mu} ((1 - e) x_A^S + x_N^S ) \approx \frac{\beta}{\gamma} ((1 - e) x_A^S + x_N^S )$,
where the approximation is valid for $\mu \ll \gamma$.
For adherers, we find in the same way $R_c^A = \frac{\beta}{\gamma} (1 - e) ((1 - e) x_A^S + x_N^S ) = (1-e) R_c^N$,

\section{The invasion boundary of adherers. \label{appendix2}}
We establish the coupled dynamics of epidemic spread and individual behavioral decisions by explicitly distinguishing between adherers and non-adherers to NPIs. 
We obtain the epidemic dynamics if there are no adherers in the population
\begin{equation}
\begin{split}
\tfrac{d}{dt}{x_N^S} =& - \beta x_N^I  x_N^S  + \delta x_N^R + \mu - \mu x_N^S \\
\tfrac{d}{dt}{x_N^I} =& +\beta x_N^I  x_N^S - \gamma x_N^I - \mu x_N^I \\
\tfrac{d}{dt}{x_N^R} =& \gamma x_N^I - \delta x_N^R - \mu x_N^R.
\end{split}
\end{equation}
There are two equilibria of this dynamics :\\
(i) A disease free equilibrium, $(x_N^S)^* = 1, $ $(x_N^I)^* = 0, $ $(x_N^R)^* = 0 $, which is stable only for \( R_0<1\);\\
(ii) An endemic equilibrium, $(x_N^S)^* = \frac{\gamma +\mu}{\beta}, $ $(x_N^I)^* =  \frac{(\beta - \gamma - \mu)(\delta + \mu)}{\beta (\gamma + \delta + \mu)}, $ $(x_N^R)^* = \frac{\gamma (\beta - \gamma - \mu)}{\beta (\gamma + \delta + \mu)} $.

We calculate the payoff in a non-adhering population
and compare it to the payoff of an individual that would switch to adherence. 
Adherence would lead to a larger payoff if
\begin{equation}
 - (1 - e) \beta \xi x_N^I - c > - \beta \xi x_N^I.
\end{equation}
From this, we get $e > \frac{c}{\beta \xi x_N^I}$.
Thus, adherers can invade a non-adhering population at the endemic equilibrium if 
\begin{equation}
e > \frac{c}{\beta \xi (x_N^I)*} = \frac{c (\gamma + \delta + \mu)}{(\beta - \gamma - \mu) ( \delta + \mu) \xi}.
\end{equation}
For \(e<  \frac{c}{\beta \xi (x_N^I)*} \), we expect no adherers in the population.


\begin{thebibliography}{10}
\expandafter\ifx\csname url\endcsname\relax
  \def\url#1{\texttt{#1}}\fi
\expandafter\ifx\csname urlprefix\endcsname\relax\def\urlprefix{URL }\fi
\expandafter\ifx\csname href\endcsname\relax
  \def\href#1#2{#2} \def\path#1{#1}\fi

\bibitem{steiner2024}
S.~Steiner, A.~Kratzel, G.~T. Barut, R.~M. Lang, E.~Aguiar~Moreira, L.~Thomann,
  J.~N. Kelly, V.~Thiel, Sars-cov-2 biology and host interactions, Nature
  Reviews Microbiology 22~(4) (2024) 206--225.

\bibitem{yamana2023}
T.~K. Yamana, M.~Galanti, S.~Pei, M.~Di~Fusco, F.~J. Angulo, M.~M. Moran,
  F.~Khan, D.~L. Swerdlow, J.~Shaman, The impact of covid-19 vaccination in the
  us: averted burden of sars-cov-2-related cases, hospitalizations and deaths,
  PLoS One 18~(4) (2023) e0275699.

\bibitem{wong2024}
B.~K.-F. Wong, N.~A. Mabbott, Systematic review and meta-analysis of covid-19
  mrna vaccine effectiveness against hospitalizations in adults, Immunotherapy
  Advances 4~(1) (2024) ltae011.

\bibitem{baker2022long}
R.~E. Baker, C.~M. Saad-Roy, S.~W. Park, J.~Farrar, C.~J.~E. Metcalf, B.~T.
  Grenfell, Long-term benefits of nonpharmaceutical interventions for endemic
  infections are shaped by respiratory pathogen dynamics, Proceedings of the
  National Academy of Sciences 119~(49) (2022) e2208895119.

\bibitem{bauch2004}
C.~T. Bauch, D.~J.~D. Earn, Vaccination and the theory of games, Proc Natl Acad
  Sci U S A 101~(36) (2004) 13391--13394.

\bibitem{perisic2009}
A.~Perisic, C.~T. Bauch, Social contact networks and disease eradicability
  under voluntary vaccination, PLoS Comput Biol 5~(2) (2009) e1000280.

\bibitem{wang2016}
Z.~Wang, C.~T.Bauch, S.~Bhattacharyya, A.~d'Onofrio, P.~Manfredi, M.~Perc,
  N.~Perra, M.~Salath{\'e}, D.~Zhao, Statistical physics of vaccination, Phys.
  Rep. 664 (2016) 1--113.

\bibitem{reluga:PLOSCB:2010}
T.~C. Reluga, Game theory of social distancing in response to an epidemic, PLoS
  Computational Biology 6~(5) (2010) e1000793.

\bibitem{flaxman2020estimating}
S.~Flaxman, S.~Mishra, A.~Gandy, H.~J.~T. Unwin, T.~A. Mellan, H.~Coupland,
  C.~Whittaker, H.~Zhu, T.~Berah, J.~W. Eaton, et~al., Estimating the effects
  of non-pharmaceutical interventions on covid-19 in europe, Nature 584~(7820)
  (2020) 257--261.

\bibitem{mendez2021systematic}
A.~Mendez-Brito, C.~El~Bcheraoui, F.~Pozo-Martin, Systematic review of
  empirical studies comparing the effectiveness of non-pharmaceutical
  interventions against covid-19, Journal of Infection 83~(3) (2021) 281--293.

\bibitem{lai2020effect}
S.~Lai, N.~W. Ruktanonchai, L.~Zhou, O.~Prosper, W.~Luo, J.~R. Floyd,
  A.~Wesolowski, M.~Santillana, C.~Zhang, X.~Du, et~al., Effect of
  non-pharmaceutical interventions to contain covid-19 in china, nature
  585~(7825) (2020) 410--413.

\bibitem{wei2023adoption}
Z.~Wei, J.~Zhuang, On the adoption of nonpharmaceutical interventions during
  the pandemic: An evolutionary game model, Risk Analysis 43~(11) (2023)
  2298--2311.

\bibitem{zamir2020non}
M.~Zamir, Z.~Shah, F.~Nadeem, A.~Memood, H.~Alrabaiah, P.~Kumam, Non
  pharmaceutical interventions for optimal control of covid-19, Computer
  methods and programs in biomedicine 196 (2020) 105642.

\bibitem{fenichel2011adaptive}
E.~P. Fenichel, C.~Castillo-Chavez, M.~G. Ceddia, G.~Chowell, P.~A.~G. Parra,
  G.~J. Hickling, G.~Holloway, R.~Horan, B.~Morin, C.~Perrings, et~al.,
  Adaptive human behavior in epidemiological models, Proceedings of the
  National Academy of Sciences 108~(15) (2011) 6306--6311.

\bibitem{bansal2007individual}
S.~Bansal, B.~T. Grenfell, L.~A. Meyers, When individual behaviour matters:
  homogeneous and network models in epidemiology, Journal of the Royal Society
  Interface 4~(16) (2007) 879--891.

\bibitem{bergstrom:PNAS:2023}
C.~T. Bergstrom, W.~P. Hanage, Human behavior and disease dynamics, Proceedings
  of the National Academy of Sciences 121~(1) (2024) e2317211120.

\bibitem{liu2022coevolution}
Y.~Liu, B.~Wu, Coevolution of vaccination behavior and perceived vaccination
  risk can lead to a stag-hunt-like game, Physical Review E 106~(3) (2022)
  034308.

\bibitem{saad-roy:PNAS:2023}
C.~M. Saad-Roy, A.~Traulsen, Dynamics in a behavioral--epidemiological model
  for individual adherence to a nonpharmaceutical intervention, Proceedings of
  the National Academy of Sciences 120~(44) (2023) e2311584120.

\bibitem{shi:MB:2024}
S.~Shi, Z.~Wang, X.~Chen, F.~Fu, Determinants of successful disease control
  through voluntary quarantine dynamics on social networks, Mathematical
  Biosciences 377 (2024) 109288.

\bibitem{bauch2005imitation}
C.~T. Bauch, Imitation dynamics predict vaccinating behaviour, Proceedings of
  the Royal Society B: Biological Sciences 272~(1573) (2005) 1669--1675.

\bibitem{bauch2013social}
C.~T. Bauch, A.~P. Galvani, Social factors in epidemiology, Science 342~(6154)
  (2013) 47--49.

\bibitem{glaubitz2023population}
A.~Glaubitz, F.~Fu, Population heterogeneity in vaccine coverage impacts
  epidemic thresholds and bifurcation dynamics, Heliyon 9~(9) (2023).

\bibitem{hu2024evolutionary}
S.~Hu, Y.~Liu, B.~Wu, Evolutionary dynamics of voluntary vaccination for
  imperfect multi-efficacy vaccines, Proceedings of the Royal Society A:
  Mathematical, Physical and Engineering Sciences 480~(2301) (2024) 20230743.

\bibitem{morsky2023impact}
B.~Morsky, F.~Magpantay, T.~Day, E.~Ak{\c{c}}ay, The impact of threshold
  decision mechanisms of collective behavior on disease spread, Proceedings of
  the National Academy of Sciences 120~(19) (2023) e2221479120.

\bibitem{karlsson:SciRep:2020}
C.-J. Karlsson, J.~Rowlett, Decisions and disease: a mechanism for the
  evolution of cooperation, Scientific Reports 10~(1) (2020) 13113.

\bibitem{tanimoto:book:2021}
J.~Tanimoto, Sociophysics approach to epidemics, Vol.~23, Springer, 2021.

\bibitem{traulsen:PNAS:2023}
A.~Traulsen, S.~A. Levin, C.~M. Saad-Roy, Individual costs and societal
  benefits of interventions during the covid-19 pandemic, Proceedings of the
  National Academy of Sciences 120~(24) (2023) e2303546120.

\bibitem{flores:npjC:2025}
L.~S. Flores, A.~d. Azevedo-Lopes, C.~M. Saad-Roy, A.~Traulsen, Seasonal social
  dilemmas, npj Complexity 2~(1) (2025) 17.

\bibitem{glaubitz:PNAS:2024}
A.~Glaubitz, F.~Fu, Social dilemma of nonpharmaceutical interventions:
  Determinants of dynamic compliance and behavioral shifts, Proceedings of the
  National Academy of Sciences 121~(50) (2024) e2407308121.

\bibitem{chen2011public}
F.~Chen, M.~Jiang, S.~Rabidoux, S.~Robinson, Public avoidance and epidemics:
  insights from an economic model, Journal of theoretical biology 278~(1)
  (2011) 107--119.

\bibitem{espinoza2022heterogeneous}
B.~Espinoza, S.~Swarup, C.~L. Barrett, M.~Marathe, Heterogeneous adaptive
  behavioral responses may increase epidemic burden, Scientific Reports 12~(1)
  (2022) 11276.

\bibitem{funk2009spread}
S.~Funk, E.~Gilad, C.~Watkins, V.~A. Jansen, The spread of awareness and its
  impact on epidemic outbreaks, Proceedings of the National Academy of Sciences
  106~(16) (2009) 6872--6877.

\bibitem{hofbauer:book:1998}
J.~Hofbauer, K.~Sigmund, Evolutionary games and population dynamics, Cambridge
  university press, 1998.

\bibitem{cressman2014replicator}
R.~Cressman, Y.~Tao, The replicator equation and other game dynamics,
  Proceedings of the National Academy of Sciences 111~(supplement\_3) (2014)
  10810--10817.

\bibitem{schuster1983replicator}
P.~Schuster, K.~Sigmund, Replicator dynamics, Journal of theoretical biology
  100~(3) (1983) 533--538.

\bibitem{heffernan:JRSI:2005}
J.~M. Heffernan, R.~J. Smith, L.~M. Wahl, Perspectives on the basic
  reproductive ratio, Journal of the Royal Society Interface 2~(4) (2005)
  281--293.

\bibitem{traulsen:NAL:2022}
A.~Traulsen, C.~S. Gokhale, S.~Shah, H.~Uecker, The {Covid-19} pandemic: Basic
  insights from basic mathematical models, Nova Acta Leopoldina 2022~(3)
  (2022).

\end{thebibliography}
\end{document}